\documentclass[12pt,a4paper]{article}
\usepackage{epsfig}
\title{Low-x contribution to the Bjorken sum rule within double logarithmic
$ln^2x$ approximation}

\begin{document}
\author{Dorota Kotlorz $^1$, Andrzej Kotlorz $^2$
\thanks{$^1$Department of Physics Ozimska 75, $^2$Department of 
Mathematics Luboszycka 3, Technical University of Opole, 
45-370 Opole, Poland, e-mail $^1$: {\tt dstrozik@po.opole.pl}}}

\pagestyle{plain}
\maketitle

\begin{abstract}
The small-$x$ contributions to the Bjorken sum rule within double logarithmic
$ln^2x$ approximation for different input parametrisations
$g_1^{NS}(x,Q_0^2)$ are presented. Analytical solutions of the evolution
equations for full and truncated moments of the unintegrated structure
function $f^{NS}(x,Q^2)$ are used. Theoretical predictions for
$\int_{0}^{0.003} g_1^{NS}(x,Q^2=10) dx$ are compared with the SMC
small-$x$ data. Rough estimation of the slope $\lambda$, controlling the
small-$x$ behaviour of $g_1^{NS}\sim x^{-\lambda}$ from the SMC data is
performed. Double logarithmic terms $\sim (\alpha_s ln^2x)^n$ become leading
when $x\rightarrow 0$ and imply the singular behaviour of 
$g_1^{NS}\sim x^{-0.4}$. This seems to be confirmed by recent experimental
SMC and HERMES data. Advantages of the unified $ln^2x$+LO DGLAP approach and
the crucial role of the running coupling
$\alpha_s=\alpha_s(Q^2/z)$ at low-$x$ are also discussed.
\end{abstract}


\section{Introduction}

The results of SIDIS (semi inclusive deep inelastic scattering) experiments
with polarised beams and targets enable the extraction of the spin dependent
quark and gluon densities. This powerful tool of studying the internal spin
structure of the nucleon allows verification of sum rules. One of them is
the Bjorken sum rule (BSR) \cite{b1}, which refers to the first moment of
the nonsinglet spin dependent structure function $g_1^{NS}(x,Q^2)$. Because
of $SU_f(2)$ flavour symmetry, BSR is regarded as exact. Thus all of
estimations of polarised parton distributions should be performed under the
assumption that the BSR is valid. Determination of the sum rules requires
knowledge of spin dependent structure functions over the entire region of
$x\in (0;1)$. The experimentally accessible $x$ range for the spin dependent
DIS is however limited ($0.7>x>0.003$ for SMC data \cite{b2}) and therefore
one should extrapolate results to $x=0$ and $x=1$. The extrapolation to
$x\rightarrow 0$, where structure functions grow strongly, is much more
important than the extrapolation to $x\rightarrow 1$, where structure
functions vanish. Assuming that the BSR is valid, one can determinate from
existing experimental data the very small-$x$ contribution ($0.003>x>0$) to
the sum rule. Theoretical analysis of the small-$x$ behaviour of 
$g_1^{NS}(x,Q^2)=g_1^{p}(x,Q^2)-g_1^{n}(x,Q^2)$ together with the broad
$x$-range measurement data allow verification of the shape of the input
parton distributions. In this way one can determinate the free parameters in
these input distributions. Experimental data confirm the theoretical
predictions of the singular small-$x$ behaviour of the polarised structure
functions. It is well known, that the low-$x$ behaviour of both unpolarised
and polarised structure functions is controlled by the double logarithmic
terms $(\alpha_s ln^2x)^n$ \cite{b3},\cite{b4}. For the unpolarised case, this
singular PQCD behaviour is however overridden by the leading Regge
contribution \cite{b5}. Therefore, the double logarithmic approximation is
very important particularly for the spin dependent structure function $g_1$.
The resummation of the $ln^2x$ terms at low $x$ goes beyond the standard LO
and NLO PQCD evolution of the parton densities. The nonsinglet polarised
structure function $g_1^{NS}$, governed by leading $\alpha_s^n ln^{2n}x$ terms, 
is a convenient function both for theoretical analysis (because of its
simplicity) and for the experimental BSR tests. The small-$x$ behaviour of
$g_1^{NS}$ implied by double logarithmic approximation has a form
$x^{-\lambda}$ with $\lambda\approx 0.4$. This or similar small-$x$ 
extrapolation of the spin dependent quark distributions have been assumed in 
recent input parametrisations e.g. in \cite{b6},\cite{b7}. In our theoretical 
analysis within $ln^2x$ approach we estimate the very small-$x$ contributions 
$\int_{0}^{x_0} g_1^{NS}(x,Q^2) dx$ and 
$\int_{x_1}^{x_2} g_1^{NS}(x,Q^2) dx$ ($x_0, x_1, x_2 \ll 1$) to
the BSR. Using analytical solutions for the full and the truncated moments of
the unintegrated structure function $f^{NS}(x,Q^2)$ \cite{b3}, \cite{b9} we
find the contributions $\int_{x_1}^{x_2} g_1^{NS}(x,Q^2) dx$ for
different input quark parametrisations: the Regge nonsingular one and the
singular ones. We compare our results with the suitable experimental SMC data 
for BSR. In the next section we recall some of the recent theoretical
developments concerning the small-$x$ behaviour of the nonsinglet polarised
structure function $g_1^{NS}$. Section 3 is devoted to the presentation of
the double logarithmic $ln^2x$ approximation, in which we calculate
analytically the full and the truncated moments of the nonsinglet function
$f^{NS}(x,Q^2)$. Section 4 contains our results for the very small-$x$
contributions to the Bjorken sum rule 
$\int_{x_1}^{x_2} g_1^{NS}(x,Q^2) dx$ ($x_1, x_2 \ll 1$). We present
our predictions using flat (nonsingular) $\sim (1-x)^3$ and singular $\sim
x^{-\lambda}$  at small-$x$ parametrisations of the input structure function
$g_1^{NS}(x,Q_0^2)$ as well. We compare our results with the SMC data for
the small-$x$ contribution to the BSR. Basing on the validity of the BSR we
roughly estimate the slope $\lambda$ controlling the small-$x$ behaviour of
$g_1^{NS}\sim x^{-\lambda}$. In Section 5 we discuss further possible
improvement of our approach. We formulate the unified equation for the
truncated moments of the unintegrated function $f(x,Q^2)$ which incorporates
$ln^2x$ resummation at low $x$ and LO DGLAP $Q^2$ evolution as well. We also
discuss the role of the running coupling effects. Finally, Section 6
contains a summary of our paper.

\section{Small-$x$ behaviour of the nonsinglet spin dependent structure
function $g_1^{NS}(x,Q^2)$}

The small value of the Bjorken parameter $x$, specifying the longitudinal
momentum fraction of a hadron carried by a parton, corresponds by definition
to the Regge limit ($x\rightarrow 0$). Therefore the small-$x$ behaviour of
structure functions can be described using the Regge pole exchange model 
\cite{b5}. In this model the spin dependent nonsinglet structure function
$g_1^{NS}=g_1^p-g_1^n$ in the low-$x$ region behave as:
\begin{equation}\label{r2.1}
g_1^{NS}(x,Q^2) = \gamma (Q^2) x^{-\alpha_{A_1}(0)}
\end{equation}
where $\alpha_{A_1}(0)$ is the intercept of the $A_1$ Regge pole trajectory, 
corresponding to the axial vector meson and lies in the limits
\begin{equation}\label{r2.2}
-0.5\leq\alpha_{A_1}(0)\leq 0
\end{equation}
This low value of the intercept (\ref{r2.2}) implies the nonsingular, flat
behaviour of the $g_1^{NS}$ function at small-$x$. The nonperturbative
contribution of the $A_1$ Regge pole is however overridden by the
perturbative QCD contributions, particularly by resummation of double
logarithmic terms $ln^2x$. In this way the Regge behaviour of the spin
dependent structure functions is unstable against the perturbative QCD
expectations, which at low-$x$ generate more singular $x$ dependence than
that implied by (\ref{r2.1})-(\ref{r2.2}). Nowadays it is well known that
the small-$x$ behaviour of the nonsinglet polarised structure function
$g_1^{NS}$ is governed by the double logarithmic terms i.e. 
$(\alpha_s ln^2x)^n$ \cite{b3},\cite{b4}. Effects of these $ln^2x$ approach go 
beyond the standard LO and even NLO $Q^2$ evolution of the spin dependent 
parton distributions and significantly modify the Regge pole model expectations 
for the structure functions. From the recent theoretical analyses of the 
low-$x$ behaviour of the $g_1^{NS}$ function \cite{b10} one can find that
resummation of the double logarithmic terms $(\alpha_s ln^2x)^n$ leads to
the singular form:
\begin{equation}\label{r2.3}
g_1^{NS}(x,Q^2) \sim x^{-\lambda}
\end{equation}
with $\lambda\approx 0.4$. This behaviour of $g_1^{NS}$ is well confirmed by
experimental data, after a low-$x$ extrapolation beyond the measured region
\cite{b2},\cite{b11},\cite{b12}.

\section{Full and truncated moments of the unintegrated structure function 
$f^{NS}(x,Q^2)$ within double logarithmic approximation}

Perturbative QCD predicts a strong  increase of the structure function
$g_1^{NS}(x,Q^2)$ with the decreasing parameter $x$ \cite{b3},\cite{b4} what is
confirmed by experimental data \cite{b2},\cite{b11},\cite{b12}. This growth is
implied by resummation of $ln^2x$ terms in the perturbative expansion. The 
double logarithmic effects come from the ladder diagram with quark and gluon
exchanges along the chain. In this approximation the unintegrated nonsinglet
structure function $f^{NS}(x,Q^2)$ satisfies the following integral
evolution equation \cite{b3}:
\begin{equation}\label{r3.1}
f^{NS}(x,Q^2)=f_0^{NS}(x)+\bar{\alpha_s}
\int\limits_x^1\frac{dz}{z}\int\limits_{Q_0^2}^{Q^2/z}
\frac{dk'^2}{k'^2}f^{NS}(\frac{x}{z},k'^2)
\end{equation}
where
\begin{equation}\label{r3.2}
\bar{\alpha_s}=\frac{2\alpha_s}{3\pi}
\end{equation}
and $f_0^{NS}(x)$ is a nonperturbative contribution which has a form:
\begin{equation}\label{r3.3}
f_0^{NS}(x)=\bar{\alpha_s}\int\limits_x^1\frac{dz}{z} g_1^{0NS}(z)
\end{equation}
The input parametrisation
\begin{equation}\label{r3.4}
g_1^{0NS}(x)=g_1^{NS}(x,Q^2=Q_0^2)
\end{equation}
in the Regge model of the low-lying trajectory $A_1$ exchange has a small-$x$
behaviour:
\begin{equation}\label{r3.5}
g_1^{0NS}(x)\sim x^{0} \div x^{0.5}
\end{equation}
More singular shape of the input function $g_1^{0NS}$ results from the
recent experimental data and PQCD analyses:
\begin{equation}\label{r3.6}
g_1^{0NS}(x)\sim x^{-\lambda}
\end{equation}
where $\lambda$ has changed during last years obtaining the values from
$0.2\div 0.3, 0.5$ \cite{b3} to recently 0.4 \cite{b10}. In our 
calculations in the next section we use different inputs $g_1^{0NS}$: the 
flat one and the singular ones at low-$x$ as well. The unintegrated 
distribution $f^{NS}(x,Q^2)$ is related to the $g_1^{NS}(x,Q^2)$ via
\begin{equation}\label{r3.7}
f^{NS}(x,Q^2)=\frac{\partial g_1^{NS}(x,Q^2)}{\partial\ln Q^2}
\end{equation}
Relation (\ref{r3.1}) implies the following equation for the truncated
Mellin moment function $\bar{f}^{NS}(x_0,n\neq 0,Q^2)$ \cite{b9}:
\begin{eqnarray}\label{r3.8}
\bar{f}^{NS}(x_0,n\neq 0,Q^2) = 
\bar{f_0}^{NS}(x_0,n) + \frac{\bar{\alpha_s}}{n}
[\int\limits_{Q_0^2}^{Q^2} \frac{dk'^2}{k'^2}\bar{f}^{NS}(x_0,n,k'^2) 
\nonumber \\
+ \int\limits_{Q^2}^{Q^2/x_0} \frac{dk'^2}{k'^2}\left(\frac{Q^2}{k'^2}\right)^n
\bar{f}^{NS}(x_0,n,k'^2)- x_0^n\int\limits_{Q_0^2}^{Q^2/x_0}\frac{dk'^2}{k'^2}
\bar{f}^{NS}(x_0,0,k'^2)]
\nonumber \\
\end{eqnarray}
where
\begin{equation}\label{r3.9}
\bar{f}^{NS}(x_0,n,Q^2) \equiv \int\limits_{x_0}^1 dx x^{n-1} f^{NS}(x,Q^2)
\end{equation}
For $x_0=0$ equation (\ref{r3.8}) reduces to that for the full Mellin moment
\begin{equation}\label{r3.10}
\bar{f}^{NS}(0,n,Q^2) \equiv \int\limits_{0}^1 dx x^{n-1} f^{NS}(x,Q^2)
\end{equation}
and in this case the analytical solution for fixed $\bar{\alpha_s}$
obtained in \cite{b3} has a form
\begin{equation}\label{r3.11}
\bar{f}^{NS}(0,n,Q^2) = 
\bar{f_0}^{NS}(0,n)\frac{n\gamma}{\bar{\alpha_s}}
\left(\frac{Q^2}{Q_0^2}\right)^{\gamma}
\end{equation}
where
\begin{equation}\label{r3.12}
\gamma = \frac{n}{2}\left[1-\sqrt{1 - (\frac{n_0}{n})^2}\right]
\end{equation}
\begin{equation}\label{r3.13}
n_0 = 2\sqrt{\bar{\alpha_s}}
\end{equation}
and $\bar{f_0}^{NS}(0,n)$ is obviously equal to
\begin{equation}\label{r3.14}
\bar{f_0}^{NS}(x_0,n) = \int\limits_{x_0}^1 dx x^{n-1} f_0^{NS}(x)
\end{equation}
at $x_0=0$. Using truncated moments approach one can avoid uncertainty from 
the unmeasurable $x\rightarrow 0$ region and also obtain important 
theoretical results incorporating perturbative QCD effects at small $x$, 
which could be verified experimentally. Truncated moments of parton 
distributions have been recently used in the LO and NLO DGLAP analysis 
\cite{b13}. In the double logarithmic approach the analytical solution of
the evolution equation (\ref{r3.8}) for fixed coupling $\bar{\alpha_s}$ has
a form \cite{b9}:
\begin{equation}\label{r3.15}
\bar{f}^{NS}(x_0,n\neq 0,Q^2) = 
\bar{f_0}^{NS}(x_0,n)\left(\frac{Q^2}{Q_0^2}\right)^{\gamma}\frac{R}{1+(R-1)x_0^n}
\end{equation}
where
\begin{equation}\label{r3.16}
R\equiv R(n,\bar{\alpha_s}) = \frac{n\gamma}{\bar{\alpha_s}}
\end{equation}
$\gamma$ is given in (\ref{r3.12}) and $\bar{f_0}^{NS}(x_0,n)$ is the
inhomogeneous term, independent on $Q^2$:
\begin{equation}\label{r3.17}
\bar{f_0}^{NS}(x_0,n) = \int\limits_{x_0}^1 dx x^{n-1} f_0^{NS}(x)=
\frac{\bar{\alpha_s}}{n}\int\limits_{x_0}^1 \frac{dx}{x} (x^n-x_0^n)
g_1^{0NS}(x)
\end{equation}
Our purpose is to calculate the truncated moments of the nonsinglet
polarised structure function $g_1^{NS}$
\begin{equation}\label{r3.18}
I(x_1,x_2,n,Q^2)\equiv\int\limits_{x_1}^{x_2} dx x^{n-1} g_1^{NS}(x,k ^2)
\end{equation}
using different input parametrisations $g_1^{0NS}$. This will allow
estimation of the small-$x$ contribution to the Bjorken sum rule and
comparison of the results with suitable experimental data. Our predictions
for $I(x_1,x_2,n,Q^2)$ will be presented in the forthcoming section.

\section{Small-$x$ contribution to the BSR within double logarithmic $ln^2x$
approximation}

Dealing with truncated moments 
$\int_{x_1}^{x_2} dx x^{n-1} g_1^{NS}(x,Q^2)$ one can avoid
uncertainties  from the experimentally unavailable regions $x\rightarrow 0$
and $x\rightarrow 1$. Particularly important is knowledge of the small-$x$
behaviour of the structure functions. In this limit $x\rightarrow 0$
$g_1^{NS}$ increases as $x^{-0.4}$ \cite{b10}, what is confirmed by a 
low-$x$ extrapolation of experimental data \cite{b2},\cite{b11},\cite{b12}. 
This extrapolation of $g_1^{NS}$ ($g_1^{p}$ and $g_1^{n}$) beyond the measured 
region of $x$ is necessary to compute its first moment $\Gamma_1$ and test 
the BSR. The BSR is a fundamental rule and must be hold as a rigorous 
prediction of QCD in the limit of the infinite momentum transfer $Q^2$:
\begin{equation}\label{r4.1}
I_{BSR} \equiv \Gamma_1^p - \Gamma_1^n =  
\int\limits_{0}^{1} dx g_1^{NS}(x,Q ^2) = \frac {1}{6}|{\frac{g_A}{g_V}}|
\end{equation}
where
\begin{equation}\label{r4.2}
\Gamma_1^p \equiv \int\limits_{0}^{1} dx g_1^{p}(x,Q ^2)
\end{equation}
\begin{equation}\label{r4.3}
\Gamma_1^n \equiv \int\limits_{0}^{1} dx g_1^{n}(x,Q ^2)
\end{equation}
and $|{\frac{g_A}{g_V}}|$ is the neutron $\beta$-decay constant
\begin{equation}\label{r4.4}
|{\frac{g_A}{g_V}}| = F + D = 1.2670
\end{equation}
Hence the BSR for the flavour symmetric sea quarks scenario
($\Delta\bar{u}=\Delta\bar{d}$) reads:
\begin{equation}\label{r4.5}
I_{BSR}(Q^2) \equiv \int\limits_{0}^{1} dx g_1^{NS}(x,Q ^2) \approx 0.211
\end{equation}
The small-$x$ contribution to the BSR has a form:
\begin{equation}\label{r4.6}
\Delta I_{BSR}(x_1,x_2,Q^2) \equiv \int\limits_{x_1}^{x_2} dx g_1^{NS}(x,Q ^2)
\end{equation}
Taking into account (\ref{r3.15})-(\ref{r3.17}) and also the relation
\begin{eqnarray}\label{r4.7}
\bar{g_1}^{NS}(x_0,n,Q^2) \equiv
\int\limits_{x_0}^1 dx x^{n-1} g_1^{NS}(x,k ^2)=
\int\limits_{x_0}^1 dx x^{n-1} g_1^{0NS}(x)
\nonumber \\
+ \int\limits_{Q_0^2}^{Q^2/x_0} \frac{dk'^2}{k'^2(1+\frac{k'^2}{Q^2})}
\bar{f}^{NS}(x_0,n,k'^2)
\nonumber \\
\end{eqnarray}
one obtains immediately:
\begin{equation}\label{r4.8}
\Delta I_{BSR}(x_1,x_2,Q^2) = I(x_1,1,1,Q ^2) - I(x_2,1,1,Q ^2)
\end{equation}
where $I(x_i,x_j,n,k ^2)$ defined in (\ref{r3.18}) in a case of $x_j=1$ has
a form:
\begin{eqnarray}\label{r4.9}
I_1(x_i,1,n,Q^2)\equiv\int\limits_{x_i}^1 dx x^{n-1} g_1^{NS}(x,k ^2)=
\bar{g_1}^{0NS}(x_i,n) 
\nonumber \\
+ B(x_i,n,Q^2)[\bar{g_1}^{0NS}(x_i,n)-x_i^n\bar{g_1}^{0NS}(x_i,0)]
\nonumber\\
\end{eqnarray}
$B(x_i,n,Q^2)$ in the right-hand side of (\ref{r4.9}) is defined as
\begin{equation}\label{r4.10}
B(x_i,n,Q^2) = \frac{\gamma (\frac{Q^2}{Q_0^2})^{\gamma}}{1+(R-1)x_i^n}
\int\limits_{\ln \frac{Q_0^2}{Q^2}}^{\ln \frac{1}{x_i}} dt 
\frac {e^{\gamma t}}{1+e^t}
\end{equation}
and
\begin{equation}\label{r4.11}
\bar{g_1}^{0NS}(x_i,n)\equiv \int\limits_{x_i}^1 dx x^{n-1} g_1^{0NS}(x)
\end{equation}
In Table I we present our results for the low-$x$ contributions to the BSR 
(\ref{r4.6}) together with $\varepsilon (x_1,x_2)$, which is defined by the
following expression:
\begin{equation}\label{r4.12}
\int\limits_{x_1}^{x_2} dx g_1^{NS}(x,Q ^2) = 
[1+\varepsilon (x_1,x_2)] \int\limits_{x_1}^{x_2} dx g_1^{0NS}(x)
\end{equation}
In the last column we give the percentage value $p [\%]$:
\begin{equation}\label{r4.13}
p = \frac{\Delta I_{BSR}(x_1,x_2,Q^2)}{I_{BSR}(Q^2)}\cdot 100\%
\end{equation}
\begin{table}[ht]
\begin{center}
\begin{tabular}{|c|c|c|c|c|}
\hline\hline
$x_1$ & $x_2$ & $\Delta I_{BSR}(x_1,x_2,10)$ & $\varepsilon (x_1,x_2)$ & p\% \\
 \hline\hline
         &                  & (1) 0.003837 & 0.5211 & 1.82  \\ \cline{3-5}
         &                  & (2) 0.018078 & 0.2242 & 8.57  \\ \cline{3-5}
   0     & $3\cdot 10^{-3}$ & (3) 0.021403 & 0.1860 & 10.14 \\ \cline{3-5}
         &                  & (4) 0.045266 & 0.0694 & 21.45 \\ \hline
         &                  & (1) 0.011682 & 0.4039 & 5.54  \\ \cline{3-5}
         &                  & (2) 0.036448 & 0.2064 & 17.27 \\ \cline{3-5}
   0     & $10^{-2}$        & (3) 0.036998 & 0.1844 & 17.53 \\ \cline{3-5}
         &                  & (4) 0.058881 & 0.0737 & 27.91 \\ \hline
         &                  & (1) 0.001344 & 0.6102 & 0.64  \\ \cline{3-5}
         &                  & (2) 0.008846 & 0.2337 & 4.19  \\ \cline{3-5}
$10^{-5}$& $10^{-3}$        & (3) 0.011391 & 0.1865 & 5.40  \\ \cline{3-5}
         &                  & (4) 0.021620 & 0.0686 & 10.25 \\ \hline
         &                  & (1) 0.011539 & 0.4010 & 5.47  \\ \cline{3-5}
         &                  & (2) 0.034050 & 0.2037 & 16.14 \\ \cline{3-5}
$10^{-4}$& $10^{-2}$        & (3) 0.032451 & 0.1840 & 15.38 \\ \cline{3-5}
         &                  & (4) 0.035964 & 0.0790 & 17.04 \\ \hline\hline
\end{tabular}
\caption{The small-$x$ contribution to the BSR (\ref{r4.6}) for different
input parametrisations (\ref{r4.14})-(\ref{r4.17}) within $ln^2x$
approximation.}
\end{center}
\end{table}
The predictions have been found for four different input
parametrisations $g_1^{0NS}(x)$, chosen at $Q_0^2=1 {\rm GeV}^2$ (1, 2, 3
inputs) or at $Q_0^2=4 {\rm GeV}^2$ (4 input):
\begin{equation}\label{r4.14}
1.~~~~~g_1^{0NS}(x) = 0.8447(1-x)^3
\end{equation}
\begin{equation}\label{r4.15}
2.~~~~~g_1^{0NS}(x) = 0.290 x^{-0.4}(1-x)^{2.5}
\end{equation}
\begin{eqnarray}\label{r4.16}
3.~~~~~g_1^{0NS}(x) = 
\frac{1}{6} x^{-0.544}[0.4949(1-x)^{2.84}(1+9.6 x^{1.23})
\nonumber \\
+ 0.204 (1-x)^{3.77}(1+14.6 x^{1.36})]
\nonumber \\
\end{eqnarray}
\begin{eqnarray}\label{r4.17}
4.~~~~~g_1^{0NS}(x) = 
\frac{1}{6}[0.1138 x^{-0.803} (1-x)^{2.403}(1+21.34 x)
\nonumber \\
+ 0.0392 x^{-0.81} (1-x)^{3.24}(1+30.8 x)]
\nonumber \\
\end{eqnarray}
Input 1 is the simple Regge form, constance as $x\rightarrow 0$; input 2
is a "toy" model, in which we have used the latest theoretical results 
concerning the small-$x$ behaviour $x^{-0.4}$ of the nonsinglet function 
$g_1^{NS}$ \cite{b10}. Finally, input 3 \cite{b14} and input 4 \cite{b15}
are parametrisations obtained lately within experimental data analysis. They 
are quite different because there are nowadays still no direct experimental 
data from the low-$x$ region. The extrapolation of $g_1^{NS}$ to very 
small-$x$ region depends strongly on the assumption (input parametrisation) 
made for this extrapolation. In Fig.1 we plot all used inputs $g_1^{0NS}(x)$ 
in the small-$x$ region $[10^{-4}\div 10^{-2}]$. In Fig.2 we present the 
small-$x$ contribution to the BSR $\Delta I_{BSR}(0,x,10)$ as a function of 
$x$ and in Fig.3 the value of $\varepsilon (0,x)$ also as a function of $x$ 
is shown. Numbers at each plot correspond to the suitable inputs $1\div 4$.
\begin{figure}[ht]
\begin{center}
\includegraphics[width=90mm]{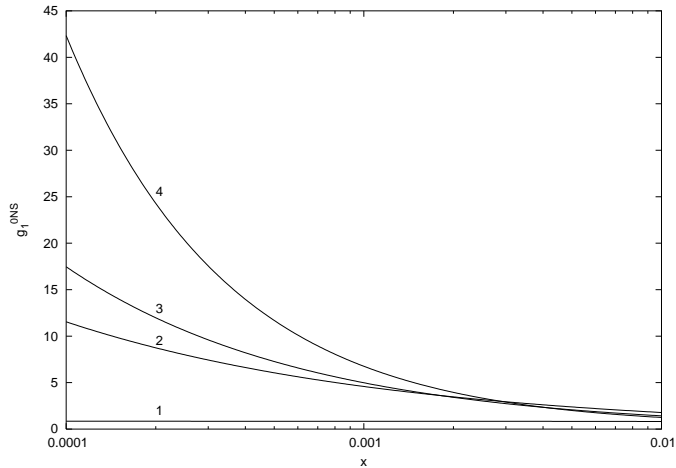}
\caption{Input parametrisations $g_1^{0NS}$ (\ref{r4.14})-(\ref{r4.17}) 
in the small-$x$ region.}
\end{center}
\end{figure}
\begin{figure}[ht]
\begin{center}
\includegraphics[width=90mm]{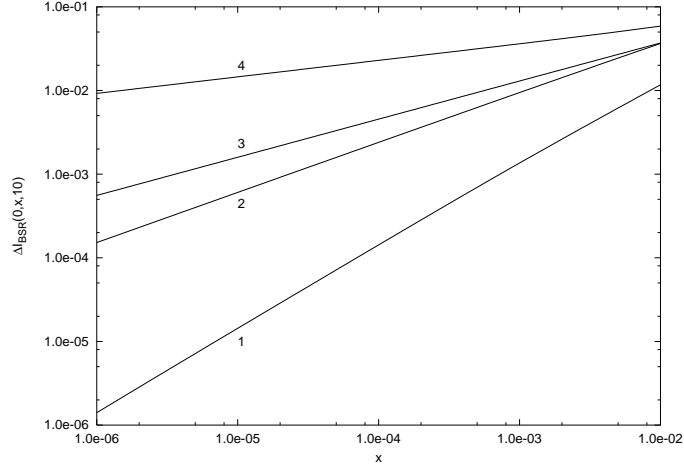}
\caption{The small-$x$ contribution to the BSR $\Delta I_{BSR}(0,x,10)$
(\ref{r4.6}) for different inputs $g_1^{0NS}$ within $ln^2x$ approach.}
\end{center}
\end{figure}
\begin{figure}[ht]
\begin{center}
\includegraphics[width=90mm]{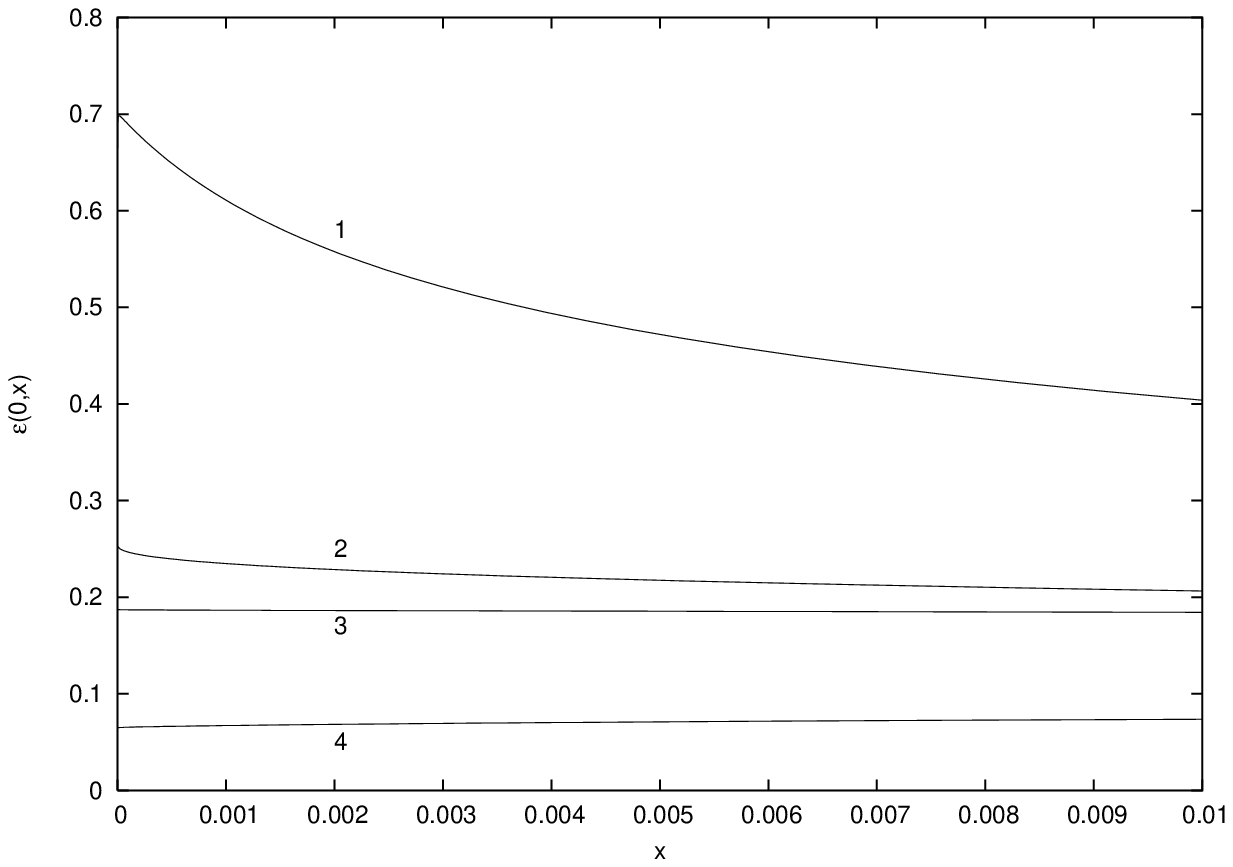}
\caption{The value of $\varepsilon (0,x)$ (\ref{r4.12}) in the low-$x$ 
region for different inputs $g_1^{0NS}$ within $ln^2x$ approach.}
\end{center}
\end{figure}
For all calculated moments $Q^2=10 {\rm GeV}^2$ and $\alpha_s = 0.18$.
From these results one can read that the low-$x$ contribution to the BSR
strongly depends on the input parametrisation $g_1^{0NS}$. For the flat
Regge form (\ref{r4.14}) $\Delta I_{BSR}(0,10^{-2},10)$ is equal to around
$5.5\%$ of the total $I_{BSR}=0.211$, while for the singular inputs
(\ref{r4.15})-(\ref{r4.17}) $18.8\%$, $17.5\%$ and $27.9\%$ respectively.
The most singular parametrisation 4 (\ref{r4.17}), predicting $x^{-0.8}$
behaviour of $g_1^{NS}$ as $x\rightarrow 0$ gives of course the largest
small-$x$ contribution to the BSR in comparison to the rest of the inputs.
The value of $\varepsilon (x_1,x_2)$, defined in (\ref{r4.12}) varies from
$0.4\div 0.7$ for the Regge input 1 to $0.07\div 0.09$ for the
"subsingular" input 4. Two similar as $x\rightarrow 0$ parametrisations 2 and
3 give $\varepsilon (x_1,x_2)$ about 0.2. It means that the double
logarithmic $ln^2x$ effects are better visible in a case of nonsingular
inputs. In a case of singular input parametrisations $g_1^{0NS}\sim
x^{-\lambda}$ ($\lambda\sim 0.4, 0.8$ or so) the growth of $g_1^{NS}$ at
small-$x$, implied by the $ln^2x$ terms resummation, is hidden behind the
singular behaviour of $g_1^{0NS}$, which survives the QCD evolution. From
the experimental SMC data \cite{b11} the low-$x$ contribution to the BSR at
$Q^2=10 {\rm GeV}^2$ is equal to
\begin{equation}\label{r4.18}
6 \int\limits_{0}^{0.003} g_1^{NS}(x,Q ^2=10) dx = 0.09\pm 0.09
\end{equation}
The above result has been obtained via an extrapolation of $g_1^{NS}$ to the
unmeasured region of $x$: $x\rightarrow 0$. Forms of the polarised quark
distributions have been fitted to SMC semi-inclusive and inclusive
asymmetries. In the fitting different parametrisations of the polarised
quark distributions \cite{b16} \cite{b17} have been used. In this way
present experimental data give only indirectly the estimation of the
small-$x$ contribution to the moments of parton distributions. The result 
(\ref{r4.18}) with a large statistical error and strongly fit-dependent
cannot be a final, crucial value. Nevertheless we would like to estimate the
exponent $\lambda$ in the low-$x$ behaviour of $g_1^{NS}\sim x^{-\lambda}$
using the above SMC result for the small-$x$ contribution to the BSR.
Assuming the validity of the BSR (\ref{r4.5}) at large $Q^2=10 {\rm GeV}^2$,
one can find:
\begin{equation}\label{r4.19}
\int\limits_{0}^{x_0} dx g_1^{NS}(x,Q ^2) = 
I_{BSR}(Q^2) - \int\limits_{x_0}^{1} dx g_1^{NS}(x,Q ^2)
\end{equation}
where $x_0$ is a very small value of the Bjorken variable. Taking into
account the small-$x$ dependence of $g_1^{NS}\sim x^{-\lambda}$ and the
experimental data for $\Delta I_{BSR}(0,0.003,10)$ one can obtain:
\begin{equation}\label{r4.20}
C \int\limits_{0}^{0.003} x^{-\lambda} dx = 0.015\pm 0.015
\end{equation}
The constant $C$ can be eliminated from a low-$x$ SMC data \cite{b11}:
\begin{equation}\label{r4.21}
C x^{-\lambda} = g_1^{n-p}(x,10)
\end{equation}
Taking different small-$x$ SMC data, we have found $\lambda =0.37$
($x=0.014$); $\lambda =0.20$ ($x=0.008$); $\lambda =0.38$ ($x=0.005$).
Comparing our predictions for $\Delta I_{BSR}(0,0.003,10)$ in Table I with
suitable SMC data, one can read that the most probably small-$x$ behaviour
of $g_1^{NS}$ is
\begin{equation}\label{r4.22}
g_1^{NS}(x,Q^2) \sim x^{-0.4}
\end{equation}
This is consistent with latest theoretical analyses \cite{b10}. The same
value of $\lambda =0.4$ was obtained in the semi-phenomenological estimation
from BSR for lower $Q^2$ \cite{b18}. Small-$x$ contribution to the Bjorken
sum rule resulting from the indirect SMC data analysis is equal to around
$7\%$ of the total value of the sum. Similar value $8.6\%$ (see Table I)
gives our QCD approach, which incorporates the singular input
parametrisation $g_1^{0NS} \sim x^{-0.4}$ according to the double
logarithmic  resummation effects. In the next point we will discuss further
possible improvement of our approach.

\section{More realistic treatment of the nonsinglet structure function
$g_1^{NS}$: $ln^2x$+LO DGLAP evolution and running $\alpha_s$ effects}

In the previous section we have presented small-$x$ contribution to the
Bjorken sum rule, calculated within double logarithmic $ln^2x$
approximation. In our approach we have used a constance (nonrunning)
$\alpha_s =0.18$. This simplification allows the analytical analysis of the
suitable evolution equations for truncated and full moments of the
unintegrated structure function $f^{NS}(x,Q^2)$. It has been however lately
proved \cite{b10}, that dealing with a very small-$x$ region one 
should use a prescription for the running coupling in a form $\alpha_s =
\alpha_s(Q^2/z)$. This parametrisation is theoretically more justified than 
$\alpha_s =\alpha_s(Q^2)$. Namely, the substitution $\alpha_s =\alpha_s(Q^2)$
is valid only for hard QCD processes, when $x\sim 1$. However the evolution
of DIS structure functions at small-$x$ needs "more running" 
$\alpha_s =\alpha_s(Q^2/z)$. Taking into account this running coupling
constant effects we obtain the following modified equation for the
$f^{NS}(x,Q^2)$:
\begin{equation}\label{r5.1}
f^{NS}(x,Q^2)=f_0^{NS}(x)+
\int\limits_x^1\frac{dz}{z}\bar{\alpha_s}(\frac{Q^2}{z})
\int\limits_{Q_0^2}^{Q^2/z}\frac{dk'^2}{k'^2}f^{NS}(\frac{x}{z},k'^2)
\end{equation}
This modification should lead to smaller values of $\bar{f}^{NS}(x_0,n,Q^2)$
and hence $\bar{g_1}^{NS}(x_0,n,Q^2)$ than for the constance $\alpha_s
=0.18$. We will discuss this problem in the next paper. Besides, our present
analysis is adequate only for a small-$x$ region, where the main role play
the double logarithmic $ln^2x$ effects. In the situation, when the present
experimental data do not cover the whole region of $x\in (0;1)$, theoretical
predictions for e.g. structure functions in the unmeasured low-$x$ region
cannot be directly verified. Thus theoretical analysis should concern both
the small-$x$ physics and the available experimentally larger-$x$ area as
well. Latest experimental SMC \cite{b2},\cite{b11} and HERMES \cite{b12} 
data provide results for the BSR from the region $0.003\leq x\leq 0.7$ and 
$0.023\leq x\leq 0.6$ respectively. In the very small-$x$ region exist only
indirect, extrapolated results with large uncertainties. Thus if one wants
to compare theoretical predictions with suitable real measurements, one has to
use an approach, which is proper for broader range of $x$ and $Q^2$. The
small-$x$ behaviour  of the nonsinglet spin dependent structure function
$g_1^{NS}(x,Q^2)$ is governed by the double logarithmic $ln^2x$ terms. This
approximation is however inaccurate for QCD analysis at larger values of $x$. 
Therefore the double logarithmic approach should be completed by LO DGLAP
$Q^2$ evolution. Unified description of the polarised structure function
$f^{NS}(x,Q^2)$ incorporating DGLAP evolution and the double logarithmic
$ln^2x$ effects at low-$x$ leads to the following equation for the truncated
moments $\bar{f}^{NS}(x_0,n,Q^2)$:
\begin{eqnarray}\label{r5.2}
\bar{f}^{NS}(x_0,n\neq 0,Q^2) = 
\bar{f_0}^{NS}(x_0,n) + \frac{\bar{\alpha_s}}{n}
[\int\limits_{Q_0^2}^{Q^2} \frac{dk'^2}{k'^2}\bar{f}^{NS}(x_0,n,k'^2) 
\nonumber \\
+ \int\limits_{Q^2}^{Q^2/x_0} \frac{dk'^2}{k'^2}\left(\frac{Q^2}{k'^2}\right)^n
\bar{f}^{NS}(x_0,n,k'^2)- x_0^n\int\limits_{Q_0^2}^{Q^2/x_0}\frac{dk'^2}{k'^2}
\bar{f}^{NS}(x_0,0,k'^2)]
\nonumber \\
+ \bar{\alpha_s}\int\limits_{Q_0^2}^{Q^2} \frac{dk'^2}{k'^2}
\sum\limits_{p=0}^{M} C_{pn} \bar{f}^{NS}(x_0,n+p,k'^2)
\nonumber \\
\end{eqnarray}
with fixed coupling constant $\alpha_s$. Matrix elements $C_{pn}$ are given
by
\begin{eqnarray}\label{r5.3}
C_{pn} = \delta_{p0}[\frac{3}{2} + \frac{1}{n(n+1)} - 2S_1(n)]
\nonumber \\
+ \sum\limits_{k=p}^{M}\frac{(-1)^p}{p!(k-p)!}[2\sum\limits_{i=1}^{n+1}
\frac{(i+k-1)!}{i!}x_0^i - \frac{(n+k-1)!}{n!} (x_0^n 
+ \frac{n+k}{n+1} x_0^{n+1})]
\nonumber \\
\end{eqnarray}
where
\begin{equation}\label{r5.4}
S_1(n) = \sum\limits_{i=1}^{n} \frac{1}{i}
\end{equation}
The unified description $ln^2x$ + LO DGLAP of the nonsinglet spin dependent
structure function $g_1^{NS}$ enables to determine the small-$x$
contribution to the Bjorken sum rule via medium- and large-$x$ contribution
of the unintegrated structure function $f^{NS}$. Because of the relation
\begin{equation}\label{r5.5}
\int\limits_{0}^{1} f^{NS}(x,Q^2) dx = 0
\end{equation}
which is discussed e.g. in \cite{b19} and \cite{b20}, one can find
immediately
\begin{equation}\label{r5.6}
\int\limits_{0}^{x_0} dx g_1^{NS}(x,Q^2) =
\int\limits_{0}^{x_0} dx g_1^{0NS}(x) -
\int\limits_{Q_0^2}^{Q^2/x_0} \frac{dk'^2}{k'^2(1+\frac{k'^2}{Q^2})}
\int\limits_{x_0}^{1} dx f^{NS}(x,k'^2)
\end{equation}
In this way, the small-$x$ part $\int_{0}^{x_0} dx g_1^{NS}(x,Q^2)$
can be expressed by the well known from experimental analysis the larger-$x$
contribution $\int_{x_0}^{1} dx f^{NS}(x,Q^2)$.
Improved picture of PQCD behaviour of the truncated moments of $g_1^{NS}$
incorporating the running coupling effects at small-$x$ and the unified
$ln^2x$ + LO DGLAP approach will be a topic of our next paper.

\section{Summary and conclusions}

In this paper we have estimated the contribution from the small-$x$ region
to the Bjorken sum rule. We have used the analytical solutions for the full 
and truncated moments of the nonsinglet polarised structure function
$g_1^{NS}(x,Q^2)$ within double logarithmic $ln^2x$ approximation. Our
predictions $\Delta I_{BSR}(x_1,x_2, Q^2)$ have been found for different input 
parametrisations $g_1^{0NS}(x,Q_0^2)$ with fixed coupling constant $\alpha_s
= 0.18$. These four parametrisations describe different small-$x$ behaviour
of $g_1^{0NS} = g_1^{0(p-n)}$ at $Q_0^2$: $g_1^{NS}\sim x^{-\lambda}$. We
found that the low-$x$ contribution to the BSR strongly depends on the input
parametrisation $g_1^{0NS}$. The percentage value $\Delta I_{BSR}(0,10^{-2},
Q^2 = 10)$ of the total BSR $\approx 0.211$ varies from 5.5 for the flat
Regge input 1 ($\lambda =0$) to almost 28 for the most singular input 4
($\lambda =0.8$). Input parametrisation 2 ($\lambda =0.4$), which incorporates
latest theoretical knowledge about small-$x$ behaviour of $g_1^{NS}$ driven
by $ln^2x$ terms, gives this $\Delta I_{BSR}$ about $17\%$ of the total BSR.
Comparing our results with the experimental SMC data one can see good
agreement, particularly  for the input 2, of the small-$x$ contribution
$0\leq x\leq 0.003$ to the BSR. However it must be emphasized, that SMC data
for the low-$x$ region suffer from large uncertainties. Using SMC data for
$g_1^{NS}$ at small-$x$ ($0.14$, $5\cdot 10^{-3}$, $8\cdot 10^{-3}$) we have
also estimated the exponent $\lambda$ which governs the low-$x$ behaviour of
$g_1^{NS}$. Thus we have obtained $\lambda =0.20\div 0.38$ with large
uncertainties. Latest theoretical investigations suggest singular small-$x$
shape of polarised structure functions: $\sim x^{-0.4}$ for the nonsinglet
case and even $\sim x^{-0.8}$ for the singlet one. Both these values are
indirectly confirmed by fitted experimental HERMES data. Basing on these
results, similar extrapolations of the spin dependent quark distributions
towards the very low-$x$  region have been assumed in several recent input
parametrisations $\Delta q(x,Q_0^2)$. Resummation of the $ln^2x$ terms
generates correctly the leading small-$x$ behaviour of the polarised
structure function but is inaccurate for larger values of $x$. In order to
have reliable theoretical predictions for the polarised structure functions
e.g. $g_1^{NS}(x,Q^2)$ and to compare them with the real and the
extrapolated recent experimental data, one should use unified theoretical
description of the evolution of the structure functions. Such unified
approach is the formalism, which contains the resummation of the $ln^2x$ and
the LO DGLAP $Q^2$ evolution as well. Besides, in each realistic analysis of 
the $Q^2$ evolution of the structure functions one should take into 
account the running coupling, where $\alpha_s \rightarrow \alpha_s(Q^2)$.
Latest theoretical investigations imply such introduction of the running
coupling, where $\alpha_s \rightarrow \alpha_s(Q^2/z)$. This is more
justified in the small-$x$ region. Combined $ln^2x$+LO DGLAP analysis together 
with taken into account the running coupling effects should give a correct 
description of the polarised structure function e.g. $g_1^{NS}$ in the whole
region of $x$. It is very important because of lack of the experimental data
from the very small-$x$ region ($x<0.003$). Agreement of the theoretical 
predictions e.g. for the BSR with real experimental data at medium and large 
$x$ may give hope, that for the very interesting small-$x$ region the 
suitable theoretical results are also reliable.

\section*{Acknowledgements}

We thank Boris Ermolaev for constructive remarks and useful comments
concerning the running coupling effects in the small-$x$ region.

\end{document}